\documentclass[12pt,twoside]{article}
\usepackage{fleqn,espcrc1}
\usepackage{graphicx}

\newcommand{\AmS}{{\protect\the\textfont2
  A\kern-.1667em\lower.5ex\hbox{M}\kern-.125emS}}
\newcommand{\Real}{\mathop{\rm Re}\nolimits}
\newcommand{\Imag}{\mathop{\rm Im}\nolimits}
\newcommand{\Rea}{\Real{a_{\eta N}}}
\newcommand{\Ima}{\Imag{a_{\eta N}}}

\title{
Photoproduction of $\eta$-mesons off light nuclei.
}

\author{N. V. Shevchenko $\rm {}^a$, V. B. Belyaev
   \address{Joint Institute  for Nuclear Research, Dubna, 141980, Russia},\\
  S. A. Rakityansky $\rm {}^b$, S. A. Sofianos
     \address{Physics Dept., University of South Africa,
         P.O. Box 392, Pretoria 0003, South Africa},
       and
  W. Sandhas
   \address{Physikalisches Institut, Universit\"{a}t Bonn, D-53115 Bonn, Germany}}

\begin{document}
\maketitle

\begin{abstract}
Photoproduction of $\eta$-mesons off deuteron is studied within
the Alt-Grassberger-Sandhas formalism for different
parameters of $\eta N$ interaction. The calculations revealed peaks
in the energy dependence of the total cross-section.
\end{abstract}

\section{INTRODUCTION}
Photoproduction of $\eta$-mesons off nuclei has attracted
essential interest during the last decades, experimentally
and theoretically as well.
Theoretical analysis of $(\gamma,\eta)$-reactions on nuclei is hampered 
by the three major problems: the unknown off-shell behavior of the
two-body $\gamma N \to \eta N$ amplitude, inaccuracies in the
description of the nuclear target as a many-body system, and
rescattering effects in the final state.
The simplest reaction is, of course, the process of coherent
$\eta$ photoproduction on deuteron. There are many theoretical
studies devoted to ($\gamma, \eta$) reactions on deuteron. Early
attempts to go beyond a simple impulse approximation 
led to very different conclusions~\cite{bib1} -- \cite{bib3} as do more recent
approaches based on the effective two-body
formulations~\cite{bib4},~\cite{bib5}. Moreover, the experimental
cross-section~\cite{bib6} of the reaction
\begin{equation}
\label{equn}
\gamma + d \to \eta + d
\end{equation}
in the near-threshold region is far above these theoretical
predictions. Therefore, a reliable description of $\eta$
photoproduction on deuteron is needed.

\section{FORMALISM}
To consider reaction~(\ref{equn}), we employ the exact
Alt-Grassberger-Sandhas (AGS) formalism modified to include the
electromagnetic interaction. 
The advantage of working with coupled equations involving the elastic
and rearrangement operators of the final state, is not only
suggested by questions of uniqueness, but also by the relevance of
rescattering effects which were found to give a significant
contribution to the corresponding amplitude~\cite{bib7}.
In the operator form, the AGS equations for the $\eta d$ system read
\begin{eqnarray}
\nonumber
 U_{11}(z) &=& \phantom{G_0^{-1}(z)+ } \,\;
 T_2(z) G_0(z) U_{21}(z) + T_3(z) G_0(z) U_{31}(z), \\
  \label{syst}
 U_{21}(z) &=& G_0^{-1}(z) + T_1(z) G_0(z) U_{11}(z) +
 T_3(z) G_0(z) U_{31}(z), \\
\nonumber
 U_{31}(z) &=& G_0^{-1}(z) + T_1(z) G_0(z) U_{11}(z) +
 T_2(z) G_0(z) U_{21}(z),
\end{eqnarray}
with $G_{0}(z)$ being the Green's operator of the
three particles involved, operator
$ T_{\alpha}(z) = t_{\alpha}(z - {\bf q}_{\alpha}^2/2 M_{\alpha}) $
is a two--body operator embedded in the three--body space;
$t_1 = t_{NN}$, $t_2, t_3 = t_{\eta N}$, and
operators $U_{ij}, i,j=1,2,3$ describe elastic scattering and 
rearrangement processes
(for more details see, for example, Ref.~\cite{bib9}). 

A photon can be introduced into this formalism by considering the
$\eta N$ and $\gamma N$ states as two different channels of the
same system. This means that we should replace the T-operator
$t_{\eta N}$  by a $2 \times 2$ matrix. It is clear, that such
replacements of the kernels of integral equations~(\ref{syst}) lead to the
corresponding solutions having a similar matrix form
\begin{equation}
\label{tmatr} 
t_{\eta N} \to \left(
\begin{array}{cc}
 t^{\gamma \gamma} & t^{\gamma \eta} \\
 t^{\eta \gamma}   & t^{\eta \eta}
\end{array}
\right) \qquad
U_{\alpha \beta} \to \left(
\begin{array}{cc}
 W_{\alpha \beta}^{\gamma \gamma} & W_{\alpha \beta}^{\gamma \eta} \\
 W_{\alpha \beta}^{\eta \gamma}   & W_{\alpha \beta}^{\eta \eta}
\end{array}
\right).
\end{equation}
Here $t^{\gamma \gamma}$ describes the Compton scattering,
$t^{\eta \gamma}$ the photoproduction process, and $t^{\eta \eta}$
the elastic $\eta N$ scattering.

It is technically more convenient to consider the reaction of
$\eta$-photoabsorption, which is inverse to reaction~(\ref{equn}). Then 
the photoproduction cross-section can be obtained by applying the
detailed balance principle.
 Therefore, we need the amplitude $W_{11}^{\gamma
\eta}$ which in the first order of electromagnetic interaction
can be written in the form
\begin{equation}
\label{solve}
 W_{11}^{\gamma \eta} \approx T_2^{\gamma \eta} G_0(z)
W_{21}^{\eta \eta} +
 T_3^{\gamma \eta} G_0(z) W_{31}^{\eta \eta}.
\end{equation}
It should be emphasized that via the operators $W_{21}^{\eta
\eta}$ and $W_{31}^{\eta \eta}$ in~(\ref{solve}) all the rescattering
effects are properly taken into account. The corresponding
transition amplitudes obey the system~(\ref{syst}) which is reduced further
to two coupled equations because of identity of the nucleons (see
Ref.~\cite{bib8}). It is customary to reduce the dimension of
these equations by representing the two-body T-operators in
separable form.

The S-wave nucleon-nucleon separable potential is adopted from
Ref.~\cite{bib11} with its parameters slightly modified to be
consistent with more recent NN data (see Ref.~\cite{bib8}). The
$\eta$-nucleon T-matrix is taken in the form
\begin{equation}
\label{tetan}
     t_{\eta N}(p',p;z)=({p'}^2+\alpha^2)^{-1}
    \frac{\lambda}{(z-E_0+i\Gamma/2)}(p^2+\alpha^2)^{-1}.
\end{equation}
The range parameter
$\alpha=3.316$ fm$^{-1}$ was determined in Ref.~\cite{bib12},
while $E_0$ and $\Gamma$ are the parameters of the $S_{11}$
resonance~\cite{bib13}.
The strength parameter $\lambda$ is chosen to reproduce the 
$\eta$-nucleon complex scattering length $a_{\eta N}$.
The value of $a_{\eta N}$ is not accurately known, different analysis
provided for it values in the range
$ 0.27\ {\rm fm}\le \Rea \le 0.98\ {\rm fm}$,
$ 0.19\ {\rm fm}\le \Ima \le 0.37\ {\rm fm}$ .
Recently, however, most of the authors agreed that 
$\Ima$ is around $0.3$ fm.

To construct a separable T-matrix of the reaction 
$\eta N \to \gamma N$, we used the results
of Ref.~\cite{bib15} where $t^{\eta \gamma}$ (which is equal to
$t^{\gamma \eta}$) was obtained as an element of a multi-channel T-matrix 
on the energy shell. For our calculations, we extended this
T-matrix off the energy shell,
using the Yamaguchi form-factors which become unit on the energy
shell. It is generally believed, that $t^{\gamma \eta}$ is
different for neutron and proton. We assumed that they have the
same functional form and differ by a constant factor
$  t_n^{\gamma \eta} = A \, t_p^{\gamma \eta}. $
A multipole analysis gives for this factor the 
following estimate~\cite{bib16}: $  A = -0.84 \pm 0.15. $

\section{RESULTS}
\begin{figure}[h]
\begin{minipage}[t]{80mm}
\includegraphics[width=70mm,height=50mm]{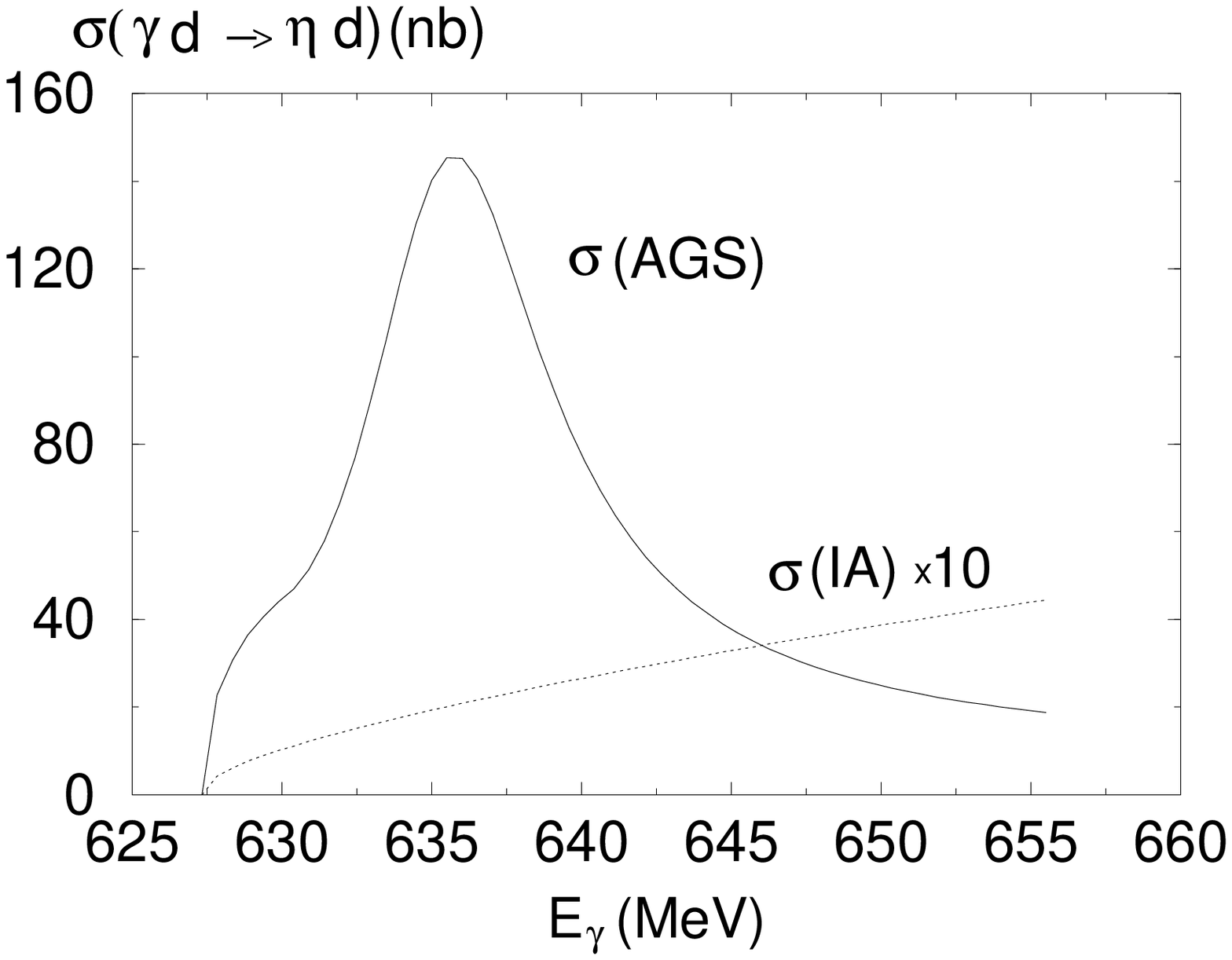}
\caption{
Total cross-section, calculated within a
rigorous few-body theory (AGS) and Impulse Approximation (IA).}
\label{fig1}
\end{minipage}
\hspace{\fill}
\begin{minipage}[t]{75mm}
\includegraphics[width=70mm,height=50mm]{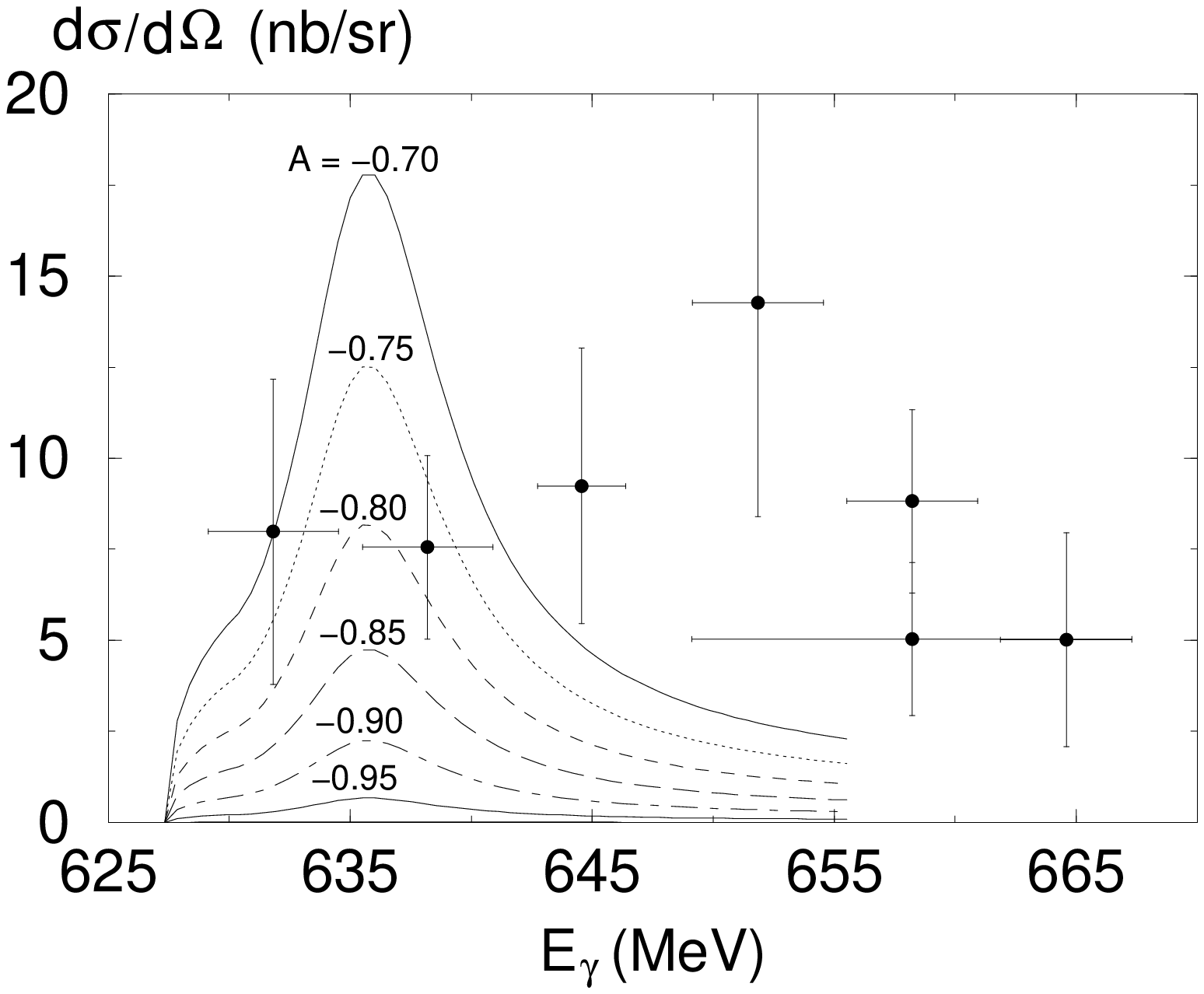}
\caption{
Differential cross-section ($\Theta_{\eta}^{cm} = 90^0$),
calculated with different choices of the ratio $A$. }
\label{fig2}
\end{minipage}
\end{figure}

\begin{figure}[htb]
\begin{minipage}[t]{80mm}
\includegraphics[width=70mm,height=50mm]{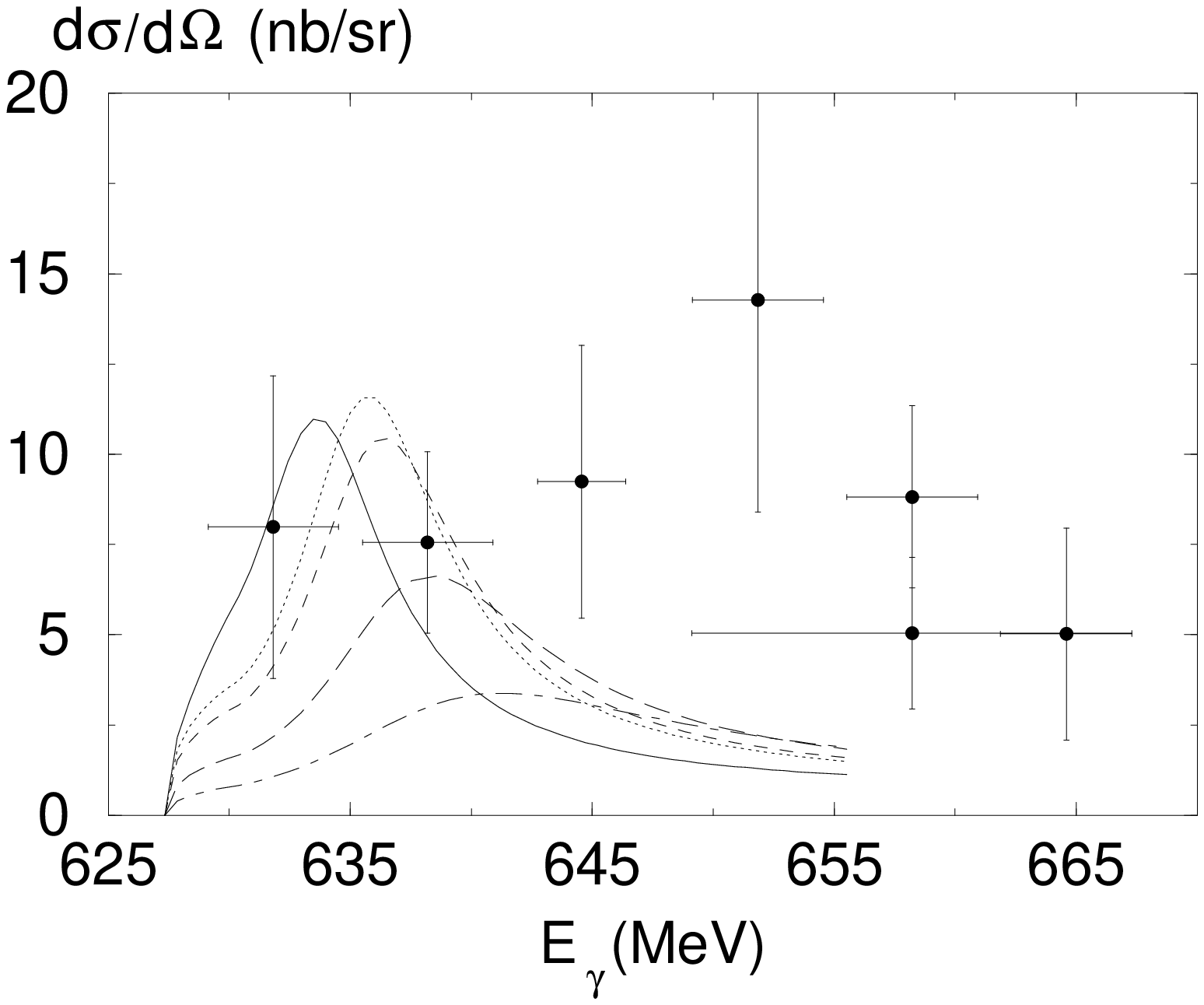}
\caption{
Differential cross-section ($\Theta_{\eta}^{cm} = 90^0$)
with different choices of $\Rea$. }
\label{fig3}
\end{minipage}
\hspace{\fill}
\begin{minipage}[t]{75mm}
\includegraphics[width=70mm,height=50mm]{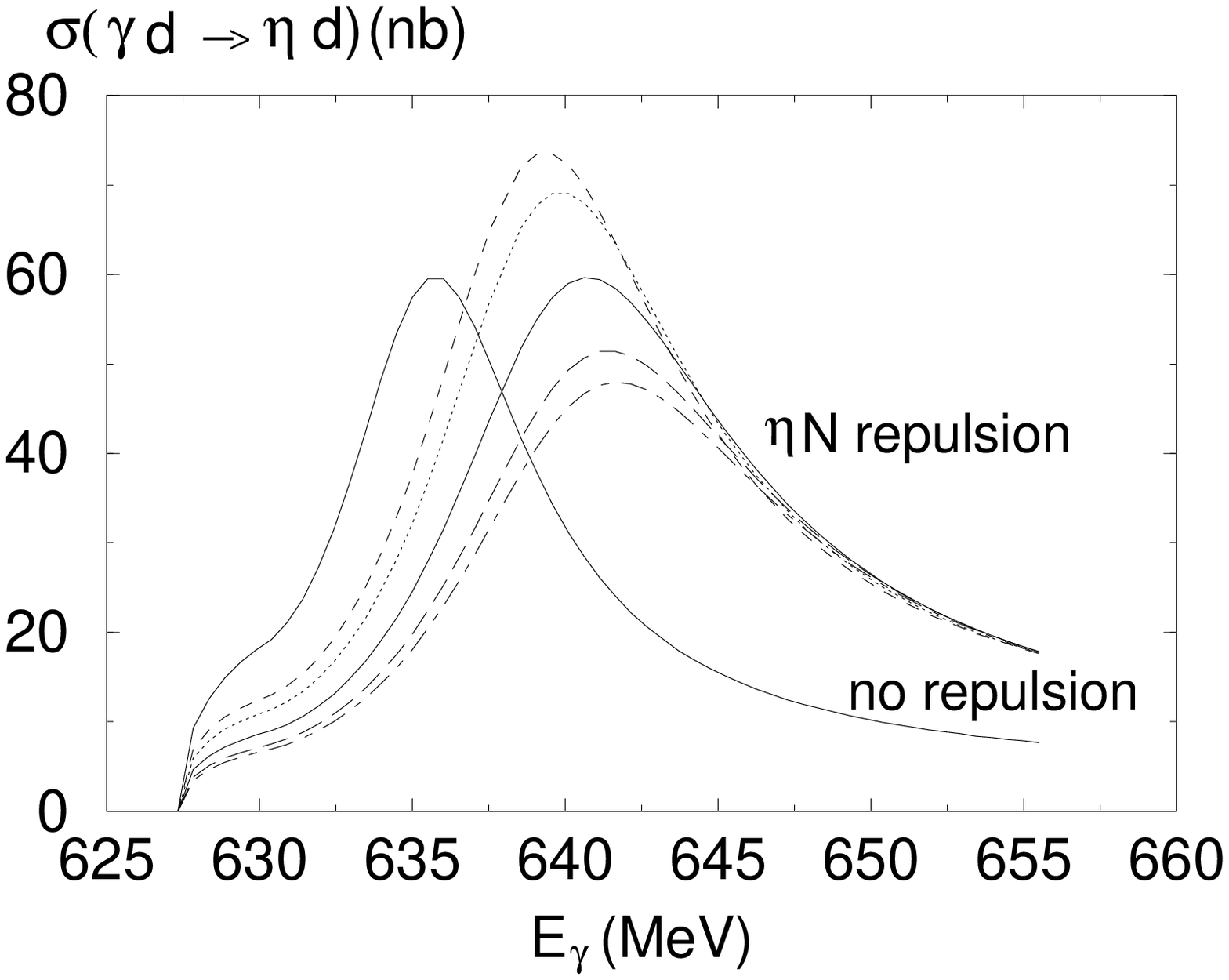}
\caption{
Shift of the resonant peak of the total cross-section 
due to the $\eta N$ repulsion.}
\label{fig4}
\end{minipage}
\end{figure}

As was expected, our calculations revealed a very strong final state
interaction in the reaction~(\ref{equn}). A comparison of the corresponding
cross-sections obtained by solving the AGS equations and by using
the Impulse Approximation (IA) is given in Fig.~\ref{fig1}, where the
IA-results are multiplied by 10.
Besides the fact that the IA-curve is generally an order of
magnitude lower, it does not show a resonant enhancement which is
clearly seen when all the rescattering and re-arrangement
processes are taken into account. In this connection, it should be
noted that experimental data~\cite{bib6} for the reaction~(\ref{equn}),
given in Figs.~\ref{fig2} and~\ref{fig3} show a pronounced enhancement of the
differential cross-section at low energies.

In order to examine the dependence of our calculations on the choice
of the parameters of the T-matrices $t^{\eta \eta}$ and $t^{\gamma
\eta}$, we did variations of $A = t_n^{\gamma \eta}/ t_p^{\gamma
\eta}$ and $\Rea$ within the corresponding
uncertainty intervals. One of the most important parameters of the
theory is the ratio of the photoproduction amplitudes for neutron and
proton ($A$). Six curves corresponding to different choices of $A$
are depicted in Fig.~\ref{fig2}. These curves were calculated with 
$a_{\eta N} = (0.75 + i 0.30)$ fm.
The experimental data are taken from Ref.~\cite{bib6}.

In Fig.~\ref{fig3}, we present the result of our calculations for five
different choices of $\Rea$, namely, 0.55 fm, 0.65 fm,
0.725 fm, 0.75 fm, and 0.85 fm.
This sequence of $\Rea$ corresponds to the upward 
sequence of the curves in the near-threshold region.
The parameter $A$ was taken to be $-0.75$.
A comparison of the curves depicted in Figs.~\ref{fig2} and~\ref{fig3} with the
corresponding experimental data shows that no agreement with the
data can be reached unless the ratio $A$ is greater than $-0.80$.

Under all variations of the parameters, however, the resonant peak
remains about 15 MeV to the left of the experimental peak. Since
this peak is due to the resonant final state interaction between
the $\eta$ meson and deuteron, we may expect that it can be
shifted to the right by introducing a repulsion into the $\eta N$
interaction.
To introduce an $\eta N$ repulsion which preserves the separable
form of the corresponding T-matrix, we used the method suggested
in Ref.~\cite{bib11} where a separable nucleon-nucleon T-matrix
includes an energy dependent factor,
$ b(E) = -\tanh{( 1 - E/E_c )}. $
This factor causes the NN phase-shift to change sign at the energy
$E_c = 0.816$ fm$^{-1}$, which is equivalent to presence of an NN
repulsion. Since the purpose of our numerical experiment was to
check if an $\eta N$ repulsion could shift the peak to the right
and there is no information about such repulsion, we used the same
function $b(E)$ and did variations of $E_c$, namely $E_c/3$,
$E_c/2$, $E_c$, $2E_c$, and $3E_c$.
The corresponding curves are shown in Fig.~\ref{fig4} where the larger
$E_c$ the lower is the curve. Parameters used here are:
$a_{\eta N} = (0.75 + i 0.30)$ fm and $A = -0.85$. 

 Therefore, comparison of our calculations with the
experimental data suggests that $A > -0.80$, $\Rea > 0.75$ fm, 
and that the $\eta N$ interaction is likely to be repulsive 
at short distances.

\vspace{5mm}

Authors would like to thank Division for Scientific Affair of NATO for
support (grant CRG LG 970110) and DFG-RFBR for financial assistance
(grant 436 RUS 113/425/1).


\begin{thebibliography}{9}
\bibitem{bib1} N. Hoshy, H. Hyuga and K. Kubodera, Nucl.Phys.
A324 (1979) 234

\bibitem{bib2} D. Halderson and A. S. Rosenthal, Nucl.Phys. A501 (1989) 856

\bibitem{bib3} Y. Zhang and D. Halderson, Phys.Rev. C45 (1992) 563

\bibitem{bib4} E. Breitmoser, H. Arenhoevel, Nucl.Phys. A612 (1997) 321

\bibitem{bib5} L. Tiator, C. Bennhold, S. S. Kamalov, Nucl.Phys. A580 
(1994) 455

\bibitem{bib6} P. Hoffman-Rothe et al., Phys.Rev.Lett. 78 (1997) 4697

\bibitem{bib7} F. Ritz and H. Arenhovel, Phys. Lett. B447 (1999) 15

\bibitem{bib8} N. V. Shevchenko et al., Phys. Rev., C58 (1998) R3055

\bibitem{bib9} V. B. Belyaev, Lectures in Few-body systems, Springer Verlag
(1990)

\bibitem{bib11} H. Garcilazo, Lett.Nuovo Cim. 28 (1980) 73

\bibitem{bib12} C. Bennhold and H. Tanabe, Nucl. Phys. A530 (1991) 625

\bibitem{bib13} Particle Data Group, Phys. Rev. D50 (1994) 1173

\bibitem{bib15} A. M. Green and S. Wycech, Phys. Rev. C60 (1999) 035208

\bibitem{bib16} N. C. Mukhopadhyay, J. F. Zhang, M. Benmerouche, Phys. Lett.
B364 (1995) 1

\end{thebibliography}
\end{document}